\documentclass{mn2e}
\usepackage{epsfig}

\usepackage{graphicx}

\newif\ifAMStwofonts
\AMStwofontstrue

\title[Constraints on the galaxy ``main sequence'' at $z>5$]
{
Constraints on the galaxy ``main sequence'' at $z>5$: the stellar mass of HDF850.1
}
\author[Serjeant \& Marchetti]
  {
{Stephen Serjeant$^{1}$ and Lucia Marchetti$^{1}$}\\  
$^1$Department of Physical Sciences, The Open University, Milton Keynes, MK7 6AA, UK\\
}
\date{Received 2013}
\pubyear{2013}

\begin{document}

\input BoxedEPS.tex
\SetEPSFDirectory{./}

\SetRokickiEPSFSpecial
\HideDisplacementBoxes

\label{firstpage}

\maketitle

\begin{abstract}
  We present rest-frame optical and near-infrared detections of one of
  the highest redshift submm-selected galaxies to date, HDF850.1.  We do
  not detect the previously proposed counterpart HDF850.1K in new deep
  $J$ and $H$-band {\it HST} WFC3 data, placing a strong limit of
  $H-K>3.8$, concluding that the $K$-band source is spurious. However,
  we detect $5.8\,\mu$m and $8\,\mu$m emission co-located with the
  submm in deblended images.  After modelling and removing the flux
  contributions from another foreground galaxy, we constrain the
  stellar mass of HDF850.1 to be $(2.5\pm1)\mu^{-1}\times10^{11}M_\odot$
  for a lensing magnification $\mu=1.9\pm0.3$, with a
  specific star formation rate of $8.5$\,Gyr$^{-1}$, faster than the
  $1-4$\,Gyr$^{-1}$ observed for $UV$-selected galaxies at this epoch.
\end{abstract}


\section{Introduction}\label{sec:introduction}

HDF850.1 was the first galaxy discovered in a blank field submm-wave
survey (Hughes et al. 1998), proving the feasibility of blank-field
sky surveys at submm wavelengths. Together with the earlier
discoveries of submm galaxies detected with the help of strong
gravitational lens magnification by foreground galaxy clusters (Smail
et al. 1997) and that of submm galaxies in other pioneering
blank-field surveys (e.g. Barger et al. 1998), this new population or
category of high-redshift ultraluminous starbursts ushered in a new
era of extragalactic survey astronomy and helped establish some of the
first observational evidence for galaxy downsizing. Despite this, even
today it remains unclear why so few $z>4$ submm galaxies have been
discovered. For example, it is not clear whether further populations
of more extreme but rarer starbursts await discovery at $z>4$ with
e.g. extremely red submm colours in {\it Herschel} SPIRE data
(e.g. Pope \& Chary 2010), or whether the median $z=2.4$ for submm
galaxies (e.g. Swinbank et al. 2004) holds at all luminosities,
marking the peak epoch of stellar mass assembly for all sufficiently
massive galaxies.

\begin{figure*}
\centering
\ForceWidth{5in}
\BoxedEPSF{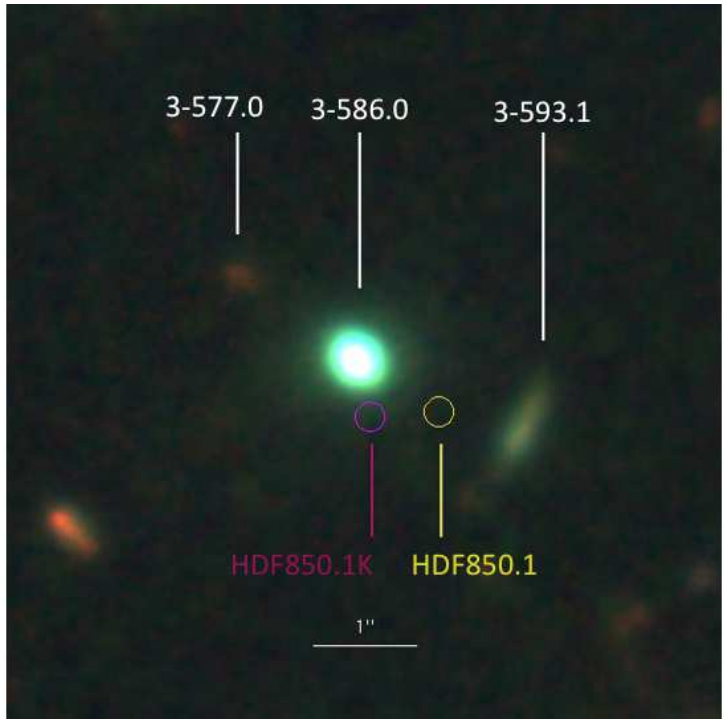}
\caption{\label{fig:hstcolour} {\it HST} composite colour image of the
  HDF850.1 system centred on 3-586.0, with $I_{{\rm F}814{\rm W}}$-band data in blue,
  $J_{{\rm F}125{\rm W}}$-band in green, and $H_{{\rm F}160{\rm
      W}}$-band in red. Named galaxies are marked. The position of
  HDF850.1K is from Dunlop et al. (2004), while HDF850.1 is shown at
  the position of the [C{\sc ii}] emission from Walter et
  al. (2012). North is up and East to the left. Note that the
  $z=1.224$ foreground elliptical 3-586.0 appears relatively blue in the near-infrared.}
\end{figure*}

\begin{figure*}
\centering
\vspace*{-2cm}
\ForceWidth{5in}
\vSlide{15cm}
\hSlide{-3cm}
\BoxedEPSF{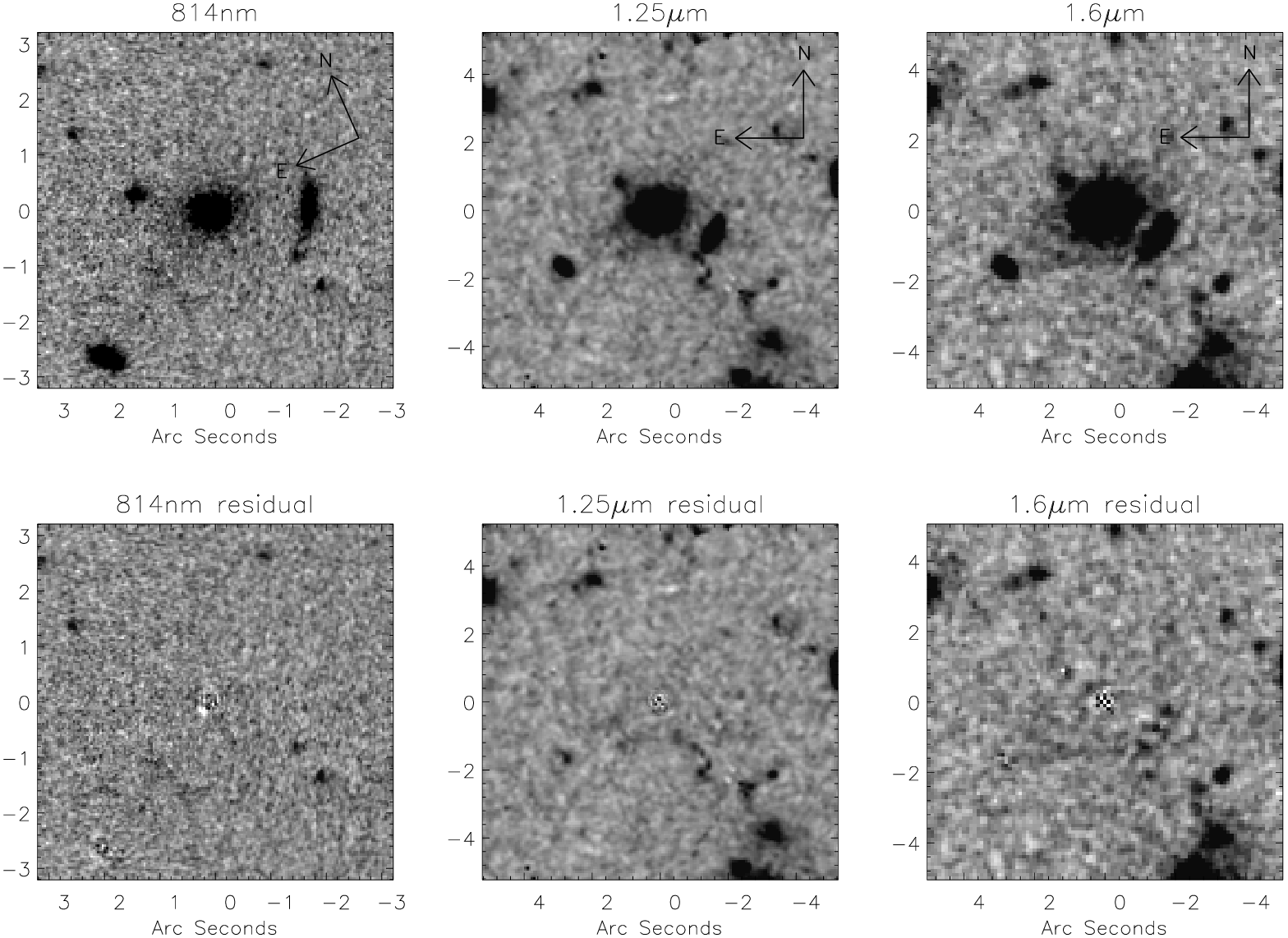}
\ForceWidth{2in}
\hSlide{11cm}
\vSlide{-6.3cm}
\BoxedEPSF{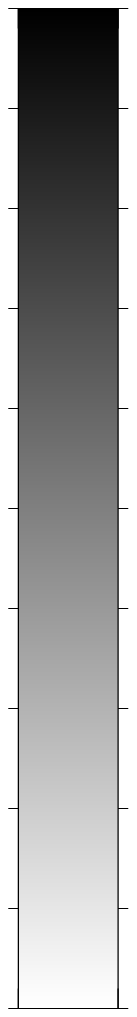}
\vspace*{-5cm}
\caption{\label{fig:hst}{\it HST} lens-subtracted imaging of HDF850.1 with
  the WFPC2 $I_{{\rm F}814{\rm W}}$-band (left), WFC3 $J_{{\rm
      F}125{\rm W}}$-band (centre) and WFC3 $H_{{\rm F}160{\rm
      W}}$-band (right). The top row shows the unsubtracted imaging,
  while the bottom row give the residuals after subtracting out
  3-577.0, 3-586.0 and 3-593.1 (Fig. \ref{fig:hstcolour}). Neither the
  submm source from Walter et al. (2012) nor the K-band source from
  Dunlop et al. (2004) are detected in the residuals. The plot limits are 
  approximately $\pm5.0\sigma$, $\pm7.4\sigma$ and $\pm4.9\sigma$ at 
  $814\,$nm, $1.25\,\mu$m and $1.6\,\mu$m respectively, where 
  $\sigma$ is the pixel noise level in each case. As with all greyscale 
  figures in this paper, the scaling is linear, and a linear greyscale 
  sidebar is shown for this figure only.
}
\end{figure*}

HDF850.1 itself nonetheless proved an unusually hard target for
multi-wavelength follow-ups, despite intensive multiwavelength
campaigns in the field (e.g. Serjeant et al. 1997, Hogg et al. 1997,
Downes et al. 1999, Aussel et al. 1999, Garrett et al. 2000, Brandt et al. 2001, 
Alexander et al. 2003, Capak et al. 2004, Morrison et al. 2010,
Conselice et al. 2011, Teplitz et al. 2011, Oliver et al. 2012, 
Guidetti et al. 2013, Teplitz et al. 2013). Dunlop et al. (2004) noted the high
likelihood association with the $z\simeq 1$ elliptical 3-586.0
($p\simeq 0.05$, Fig.\,\ref{fig:hstcolour}), but also that this
association disagreed with an IRAM $1.3$\,mm interferometric position,
The proximity of the foreground elliptical makes gravitational lensing a
significant consideration. Dunlop et al. performed careful subtraction
of the foreground elliptical in {\it Subaru} $K$-band data, finding a
faint $K\simeq23.5$ counterpart, tentatively also detected in {\it
  HST} NICMOS $H$-band data with $H-K=1.4\pm0.35$. However it
transpired that the position of this $K$-band source did not agree
with subsequent SMA $345$\,GHz imaging by Cowie et al. (2009), nor
with tentative VLA detections. Cowie et al. further state that
HDF850.1 therefore ``has no detectable optical or NIR light'', and
their submm/radio photometric redshift estimate placed this objects at
$z=4.1\pm0.5$ making it potentially one of the highest redshift submm
galaxies. 

The advent of the {\it Herschel} Space Observatory then brought many
major milestones in the study of high-$z$ submm galaxies: SPIRE's very
high survey mapping efficiency (Griffin et al. 2010) made possible
many successful and much larger blank-field surveys (e.g. Eales et
al. 2010, Oliver et al. 2012); millimetric and submm spectroscopy of
these bright submm galaxies easily yielded redshifts directly from CO
lines, without recourse to multi-wavelength identifications (e.g. Lupu
et al. 2012, Frayer et al. 2011); and the steep submm source counts
were confirmed to generate a strong gravitational lens magnification
bias, demonstrating that submm surveys are extremely efficient at
finding strong gravitational lenses (e.g. Negrello et al. 2010).

Following this, and over a decade after the original discovery of
HDF850.1, Walter et al. (2012) made a breakthrough CO and [C{\sc ii}]
redshift determination of HDF850.1. This placed the system at
$z=5.183$, making it one of the highest confirmed redshift of any submm
galaxy together with, e.g., the $z=5.2$ lensed submm galaxy identified
by Combes et al. 2012 and  
the $z=6.34$ starburst discovered by Riechers et. al 2013. 
Remarkably HDF850.1 remained nonetheless undetected in all
rest-frame ultraviolet, optical and near-infrared data, making it
appear completely obscured despite the moderate surface density of its
star formation ($\sim35$\,$M_\odot$\,yr$^{-1}$\,kpc$^{-2}$,
uncertainty $\sim50$\%; Walter et al. 2012).

In this paper we use new HST WFC3 archival near-infrared data, plus a
careful re-analysis of the archival {\it Spitzer} data, to reassess
these claimed optical/near-infrared non-detections as well as the
claimed magnification factors in this lensed system; our key results
are the first rest-frame optical/near-infrared detections of
HDF850.1. Section \ref{sec:observations} presents our observational
data. The photometric data and the gravitational lens system as a
whole are modelled in section \ref{sec:modelling}. In section
\ref{sec:discussion} we discuss the implications and context of our
results, and we conclude in section \ref{sec:conclusions}.  Throughout
this paper, we refer to the submm source as HDF850.1 and the $K$-band
source from Dunlop et al. (2004) as HDF850.1K.  We adopt a concordance
cosmology, with a Hubble constant $H_0=72$\,km\,s$^{-1}$\,Mpc$^{-1}$
and density parameters $\Omega_{\rm M}=0.3$, $\Omega_\Lambda=0.7$.

\begin{table*}
\begin{tabular}{lllll}
          & $J_{\rm F125W}$ & $H_{\rm F160W}$ & $5.8\,\mu$m/$\mu$Jy & $8\,\mu$m/$\mu$Jy\\
\hline
3-586.0   & $21.24\pm0.05$ & $20.34\pm0.05$ & $8.25\pm0.40$  & $6.88\pm0.49$ \\
HDF850.1  & $<28.2$ & $<27.3$  & $2.72\pm0.53$  & $5.93\pm0.69$\\
HDF850.1K & $<28.5$ & $<27.3$  & --  & -- \\
3-593.1   & $23.62\pm0.15$  & $23.13\pm0.15$  & $2.97\pm0.39$  & $2.87\pm0.49$ \\
\end{tabular}
\caption{\label{tab:photometry}
  New photometry of the four nearest galaxies in the HDF850.1
  field. Magnitudes are in the
  Vega system. Upper limits are $2\sigma$ in an $0.4''$ diameter aperture,
  corrected to total flux assuming a point source; $J$- and $H$-band
  detections use a $3''$ diameter aperture. The 3-583.0 photometric detections on
  the $J$- and $H$-band images are
  consistent within $1\sigma$ with photometry of the {\sc galfit}  model image; the {\sc galfit} 
  image measurements are quoted here, since they exclude neighbouring
  galaxies. The 3-593.1 $J$- and $H$-band
  measurements were made on images with the {\sc galfit}  3-586.0 model
  subtracted. IRAC fluxes are from
  the point source fitting discussed in the text. 
}
\end{table*}

\section{Observational data}\label{sec:observations}
\subsection{{\it HST} near-infrared imaging}\label{sec:hst}
The GOODS-N field, among other fields, has recently been observed with
the Wide Field Camera 3 (WFC3) on the {\it HST} under the CANDELS
survey (Grogin et al. 2011, Koekemoer et al. 2011; program 12443, PI:
Faber). We performed a 
noise-weighted coadd of all available pipeline-processed data on
HDF850.1 in the F125W ($J$-band) and F160W ($H$-band) filters and
registered the coadded images to the HDF-N frame. The total exposure
time is $10.0$ hours in F160W and $8.9$ hours in F125W, and the data
were taken between 31 March 2012 and 7 November 2012.

We modelled the foreground lensing galaxy, 3-586.0, with the {\sc
  galfit} package (Peng et al. 2010a). Our approach in modelling the
lens is to use the {\it smallest} number of components possible,
i.e. keeping the model as simple as possible. The objective is a purely
phenomenological description of this galaxy for the purposes of
subtracting its flux. At $814\,$nm, the lens is modelled as the sum of
two S\'{e}rsic profiles, though allowing five Fourier distortions of
the smaller profile. Galaxies 3-577.0 and 3-593.1 were additionally
modelled, the latter incorporating a bending mode. At $1.25\,\mu$m and
$1.6\mu$m the Fourier distortions to the smaller S\'{e}rsic profile
were not found to be necessary but an addition central point spread
function was incorporated.

Fig.\,\ref{fig:hst} shows the lens-subtracted $J$-band and
$H$-band data. The tentative $H$-band counterpart seen in shallower
NICMOS data by Dunlop et al. (2004) is not seen in this new data. To
quantify the constraint, we performed $0.4$'' diameter aperture
photometry (which we then aperture corrected to total flux assuming a
point source) at locations at approximately the same radial distance
from the lens, then calculated the standard deviation of these
measurements. The $2\sigma$ upper limits at the location of the [C{\sc
  ii}] emission is $J<28.2$ and $H<27.3$, and at the location of
HDF850.1K the limits are $J<28.8$ and $H<27.3$.  Our $H$-band
photometry of the foreground lensing galaxy 3-586.0 (table
\ref{tab:photometry}) is consistent with the Dunlop et al. (2004)
$H$-band measurement of $20.40\pm0.05$. Table \ref{tab:photometry}
lists our new photometry for the galaxies in this system. 
Our $H-K>3.8$ colour for HDF850.1K (equivalent to $H_{\rm
  AB}-K_{\rm AB}>3.3$) contrasts with the $H-K=1.4$ measurement in
Dunlop et al. (2004) from their tentative NICMOS detection.

\begin{figure}
\centering
\ForceWidth{3.3in}
\hSlide{0.5cm}
\BoxedEPSF{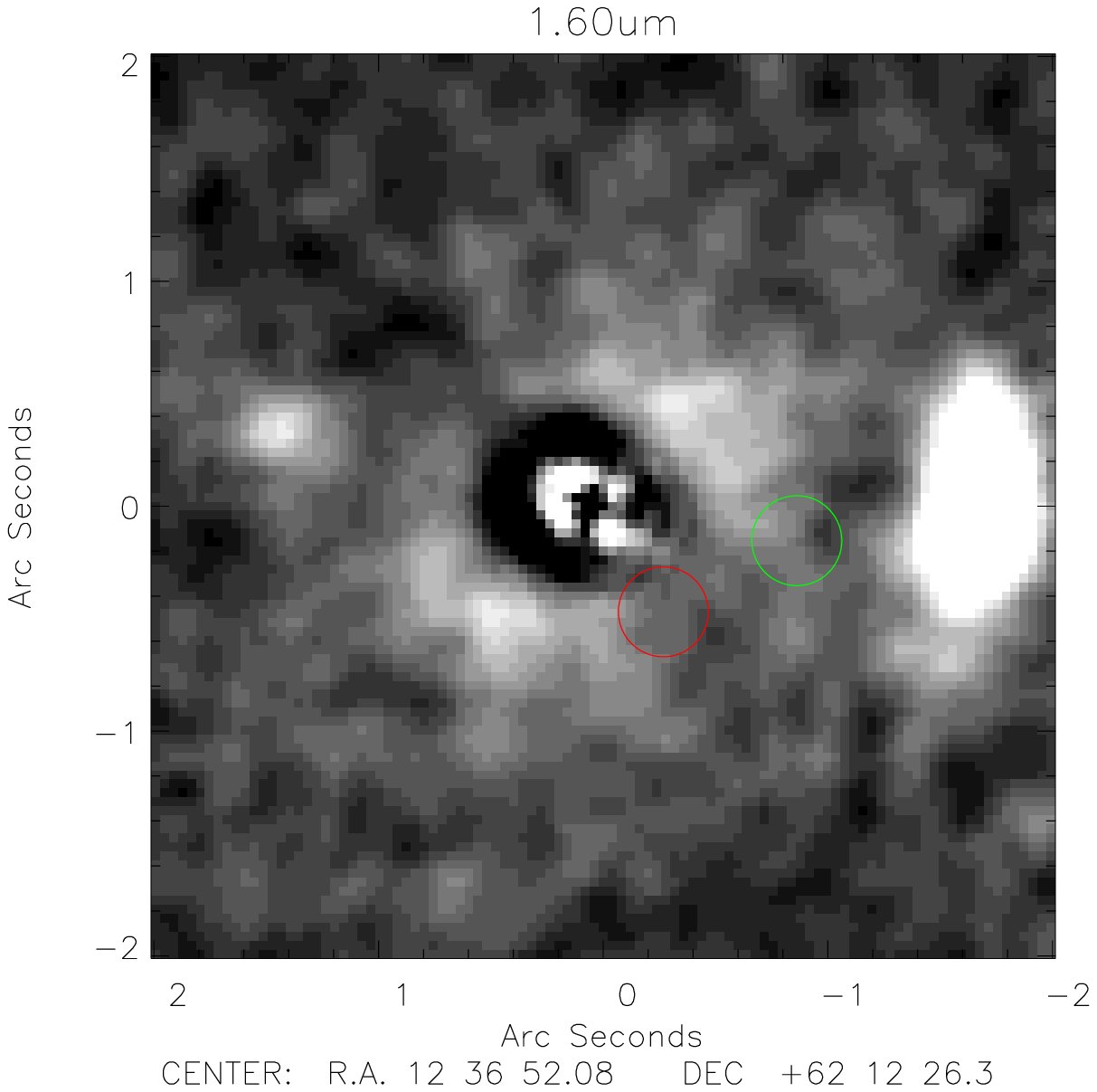}
\ForceWidth{3.3in}
\hSlide{0.5cm}
\BoxedEPSF{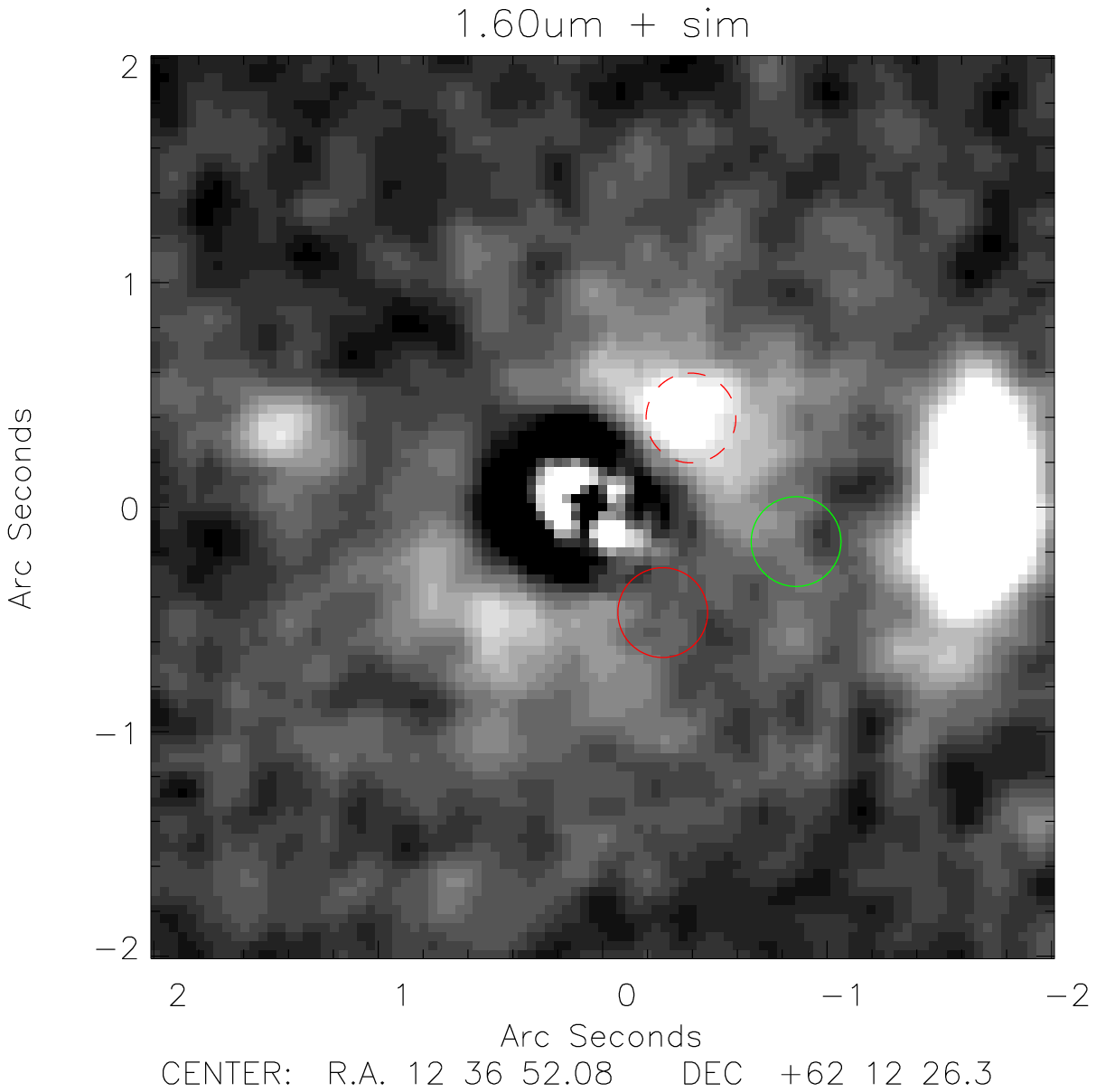}
\caption{\label{fig:hstplussim} {\it HST} lens-subtracted WFC3
  $H_{{\rm F}160{\rm W}}$-band data (top) with a {\it maximally
    simple} lens model of a single S\'{e}rsic profile. The image
  pixels are oversampled in these images for clarity of presentation,
  though this accentuates residuals in the image centre, due to the
  sensitivity to the modelling of the PSF on scales significantly
  smaller than a detector pixel. Note the faint diffuse residual
  flux. In the bottom panel, a simulated source has been inserted at
  the location of the dashed red circle, with the reported $H$-band
  flux from Dunlop et al. (2004). The green circle marks the position
  of the submm [C{\sc ii}] emission in Walter et al. (2012), while the
  red circle marks the location of the proposed $K$-band
  identification from Dunlop et al. (2004). If there were an $H$-band
  source at the location of HDF850.1K with the previously reported
  $H$-band flux, it would have been easily detectable in this image,
  even in this minimally-complex subtraction, so the non-detection is
  unlikely to be an artefact of our lens galaxy modelling.}
\end{figure}

As a further simple test of the lens subtraction, we attempted
inserting a simulated source at the same radial distance from the lens
as HDF850.1K, using a single S\'{e}rsic profile lens model in order to
demonstrate the robustness of the lens modelling. Accordingly, we
rescaled the $H$-band image by a factor of $1/63$, equivalent to the
reported $H$-band flux ratio of the lens and HDF850.1K in Dunlop et
al. (2004), offset the rescaled image, added it to the $H$-band image,
and finally re-performed the lens subtraction. The offset was
performed in a different direction to the HDF850.1K-lens offset, but
with the same magnitude. The result is shown in
Fig.\,\ref{fig:hstplussim}. Note the clear detection of the simulated
counterpart. We conclude that if the tentative $H$-band NICMOS
detection were real, it should easily have been reproduced in our
data.

\subsection{{\it Spitzer} near-infrared imaging}\label{sec:spitzer}
HDF850.1 was observed as part of the Great Observatories Origins Deep
Program (GOODS) in 2004 (Dickinson et al. 2003) using the Infrared
Array Camera (IRAC) and the Multiband Imaging Photometer for SIRTF
(MIPS). In this paper we will deal only with the IRAC GOODS data,
because the MIPS resolution is too poor to resolve the system.  The
{\it Spitzer} IRAC data were taken at two epochs. The fortunate
central location of HDF850.1 within GOODS led to its being observed in
both epochs. We registered the data from both epochs separately to
galaxies local to HDF850.1 in the $I_{{\rm F}814{\rm W}}$-band {\it
  HST} WFPC2 data. 

\begin{figure*}
\centering
\ForceWidth{3.3in}
\BoxedEPSF{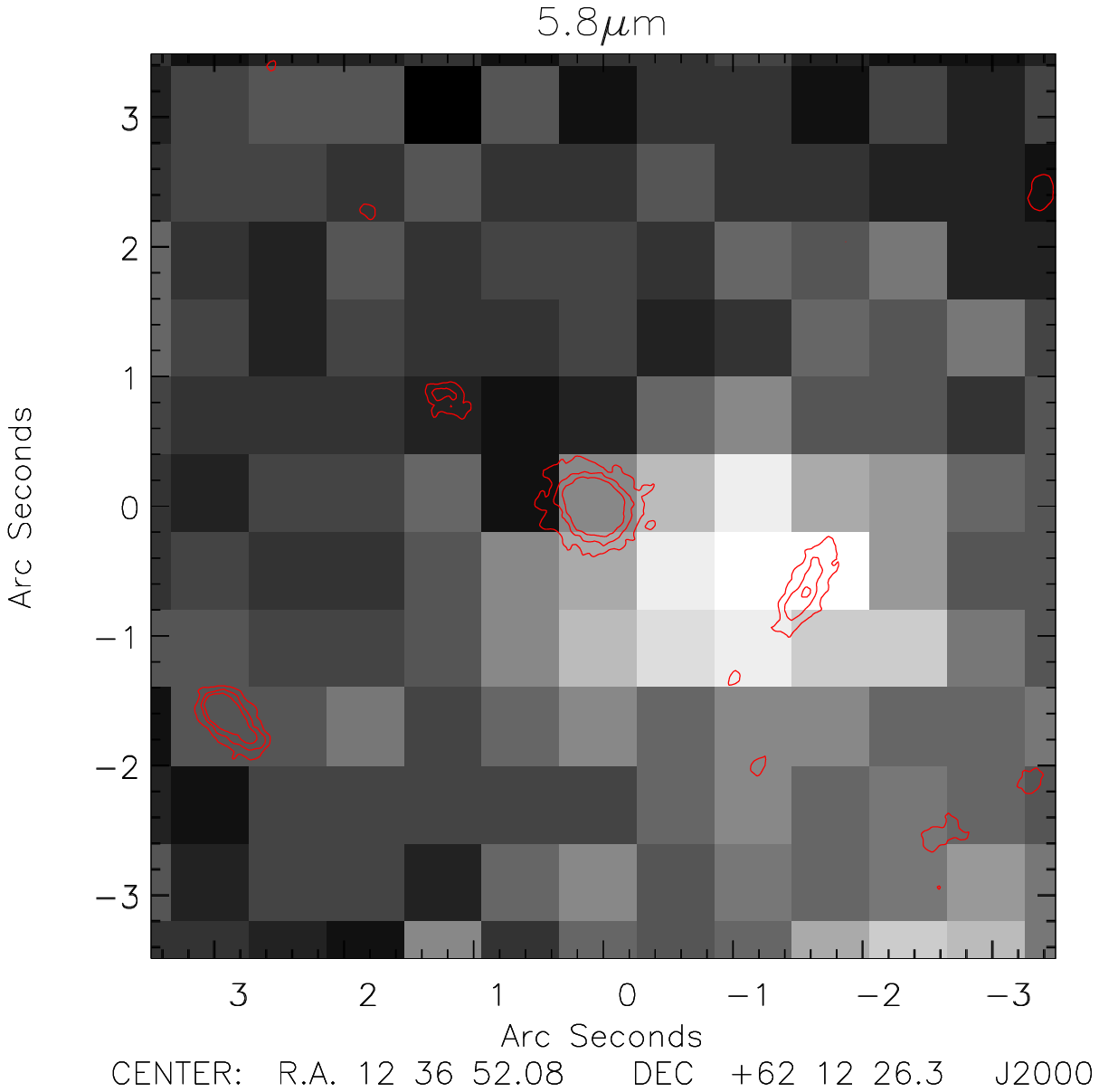}
\ForceWidth{3.3in}
\BoxedEPSF{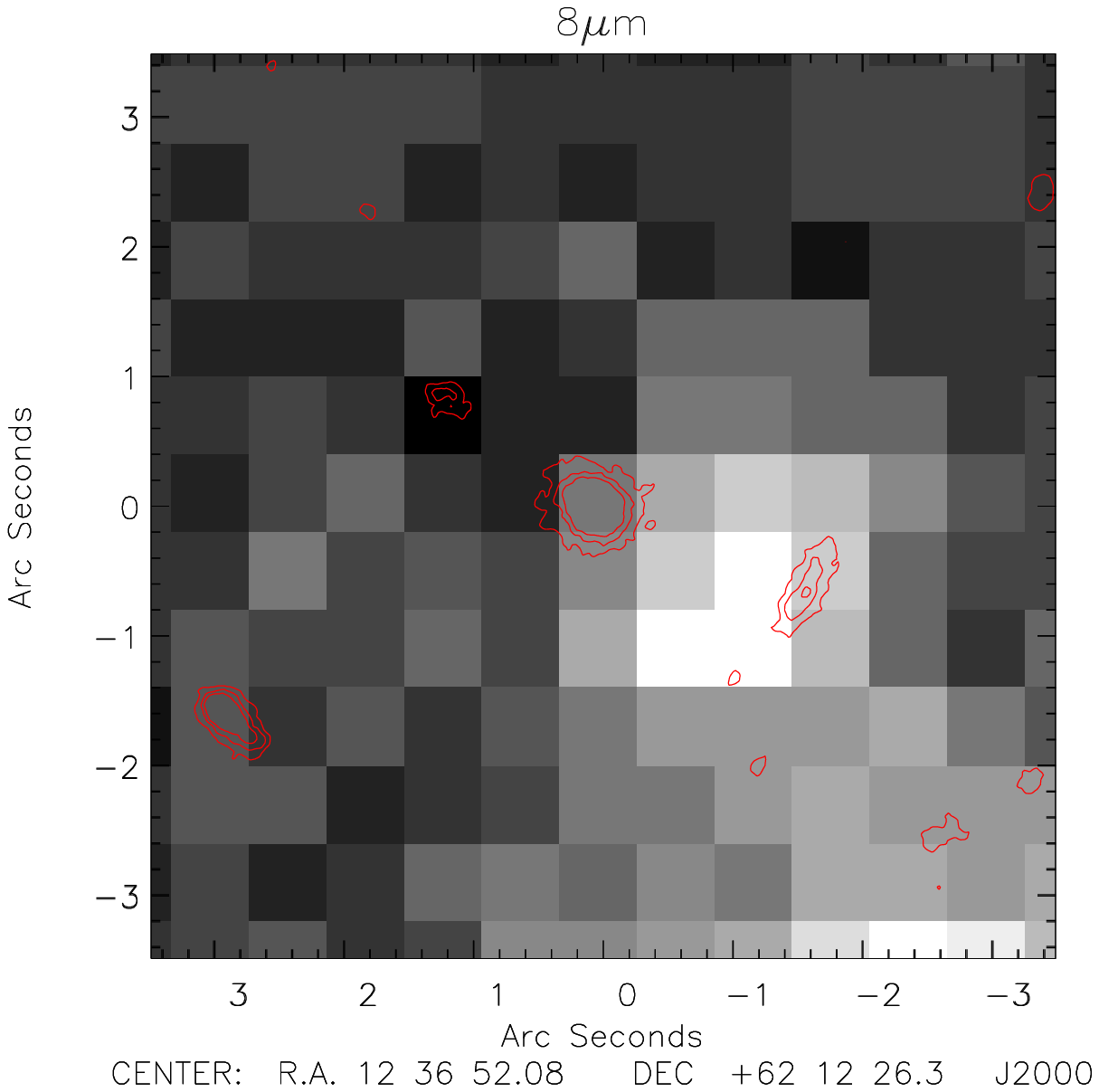}
\caption{\label{fig:ch3ch4} {\it Spitzer} lens-subtracted $5.8\,\mu$m
  and $8\,\mu$m images. North is up and East to the left. The
  foreground elliptical 3-586.0 was subtracted separately in the two
  epochs of {\it Spitzer} data, which were then combined. The red
  contours trace the $I_{{\rm F}814{\rm W}}$-band {\it HST} WFPC2
  data. Note that the {\it Spitzer} residual is not consistent with
  emission solely from the location of the foreground galaxy 3-593.1
  (Fig.\,\ref{fig:hstcolour}). The $5.8\,\mu$m image is scaled from 
  approximately $-2.6\sigma$ to $8.5\sigma$, while the $8\,\mu$m image 
  is scaled from approximately $-3.9\sigma$ to $12.9\sigma$ where in 
  each case $\sigma$ is the pixel noise level.}
\end{figure*}

The close proximity of HDF850.1 to several foreground galaxies
(Fig.\,\ref{fig:hstcolour}) complicates the {\it Spitzer} photometry
of HDF850.1 (e.g. Cowie et al. 2009). Further complicating the
analysis is the anisotropic point spread function, particularly at the
shorter wavelengths. The effective radius of the foreground lens
3-586.0 is much smaller than that of an IRAC pixel, so opted to treat
the lens as a point source for {\it Spitzer}, in keeping with our
philosophy to use models with only just sufficient complexity to
characterise the lens. We used a $5\times5$ oversampled point spread
function in each epoch to construct the model profile of the
foreground lens at $5.8\,\mu$m and $8\,\mu$m. The lens-subtracted
images were rebinned to the same scale and coadded, and the coadded
$5.8\,\mu$m and $8\,\mu$m data is shown in Fig.\,\ref{fig:ch3ch4}. As
this adequately subtracted the lens we did not attempt more elaborate
lens modelling. An excess emission to the South-West of 3-586.0 is
clearly visible in both channels, as well as in the separate epochs
(not shown). Furthermore, the centroid of this emission is clearly not
consistent with flux solely from the neighbouring foreground galaxy
3-593.1. In section \ref{sec:seds} we will model the SED of 3-593.1 to
assess the likely contribution from this galaxy to the {\it Spitzer}
flux.  Fig.\,\ref{fig:ch3ch4} makes it immediately clear that the
cause of the previous non-detection of HDF850.1 is not heavy
obscuration in the submm galaxy (c.f. Walter et al. 2012), but rather
the blending from foreground systems. (HDF850.1 may nonetheless still
be heavily obscured.)

The positions of all the galaxies are known, and given that the data
is already inconsistent with flux at only 3-586.0 and 3-593.1, we
tried a three-component fit to the $5.8\,\mu$m and $8\,\mu$m images,
with point sources fixed at the locations of 3-586.0, 3-593.1 and the
submm emission. As before, the PSF subtraction was performed
separately on the two epochs, though the epochs were constrained to
have the same fluxes and source positions in the modelling. We
detected the submm source at $8.3\sigma$ at $8\,\mu$m, and $5.1\sigma$
at $5.8\,\mu$m. The photometric measurements are given in table
\ref{tab:photometry}. The measured fluxes of the submm source are
anticorrelated with the foreground galaxies; covariance matrices are
given in tables \ref{tab:covar5} and \ref{tab:covar8}. At both
wavelengths and both epochs, the residuals are consistent with blank
fields. It is not possible with this data to separate the IRAC fluxes
of HDF850.1 and HDF850.1K; we will return to the reality of HDF850.1K
in section \ref{sec:discussion}. 

\begin{figure*}
\centering
\ForceWidth{6in}
\BoxedEPSF{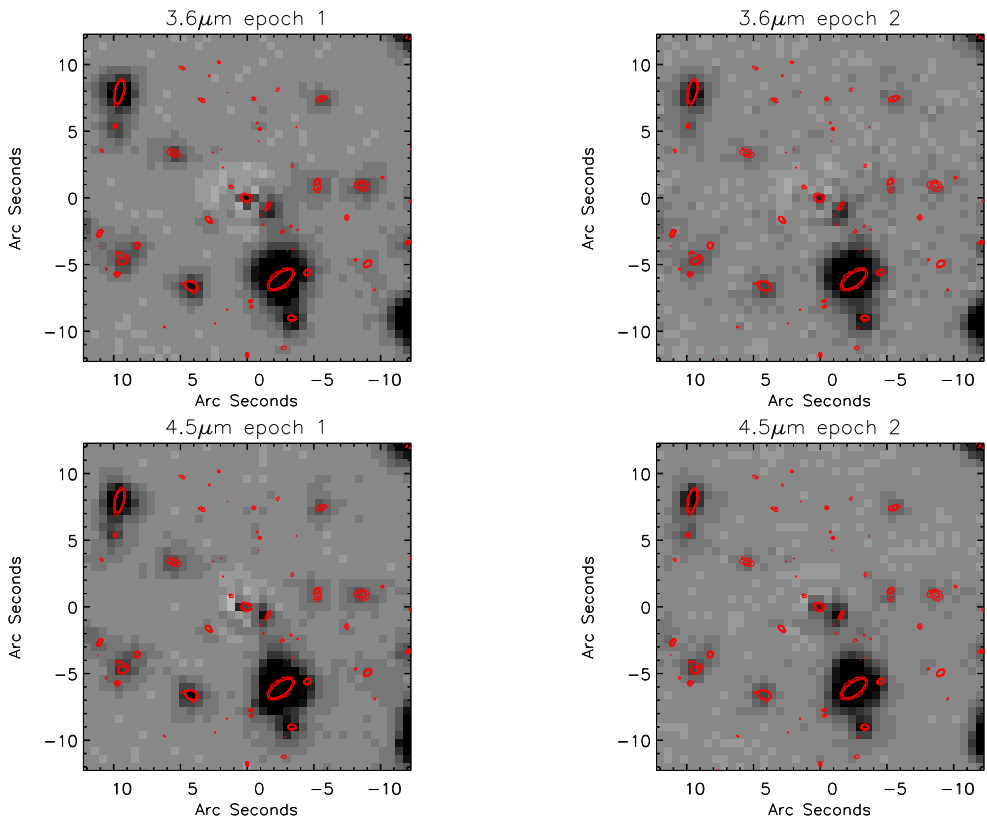}
\caption{\label{fig:ch1ch2} {\it Spitzer} lens-subtracted $3.6\,\mu$m
  and $4.5\,\mu$m images. The extended component of the foreground
  elliptical 3-586.0 (only slightly extended at this pixel scale) has
  been subtracted, but the other components have been left. There is
  evidence for a central residual point source in the lens.  The red
  contours trace the $I_{{\rm F}814{\rm W}}$-band {\it HST} WFPC2
  data. The greyscales are scaled show approximately $\pm28\sigma$ 
  where $\sigma$ is the pixel noise level.}
\end{figure*}

The lens subtraction at $3.6\,\mu$m and $4.5\,\mu$m is much more
complicated due to the highly anisotropic point spread function that
varies across the detectors. Furthermore, there is a less favourable
contrast ratio of the background sources against the foreground
lens. We nonetheless attempted to model the net PSF using the {\sc
  TinyTim} software (Krist et al. 2011), using the GOODS-N {\it
  Spitzer} coverage maps to estimate the distribution of detector
pixel locations for the lens at each epoch. Reasonably acceptable fits
were found to be possible using the $1.6\,\mu$m {\sc galfit} solution
as a starting point. Fig.\,\ref{fig:ch1ch2} shows the two epochs at
$3.6\,\mu$m and $4.5\,\mu$m. We found that the residuals depend
sensitively on the assumed point spread function and the lens model
components, so for the purposes of constraining the flux at the
position of the submm galaxy these data are less useful than the
longer wavelength IRAC channels, despite the larger PSFs at longer
wavelengths, because the longer-wavelength PSFs are less
asymmetrical. We therefore conservatively opt not to use the
$3.6\,\mu$m and $4.5\,\mu$m data to constrain the fluxes of the submm
galaxy. Nevertheless, the approximate {\it combined} fluxes of HDF850.1 and
3-593.1 of $\sim2.6\,\mu$Jy and $\sim2.9\,\mu$Jy at $3.6\,\mu$m and
$4.5\,\mu$m respectively are consistent with our SED fits to the
individual galaxies discussed below, which yield totals of
$1.7\,\mu$Jy and $3.0\,\mu$Jy respectively.

\begin{table}
\begin{tabular}{l|lll}
  & 3-586.0 & HDF850.1 & 3-593.1\\
\hline
3-586.0 & 0.0023 & -0.0066 & 0.0018\\
HDF850.1 & -0.0066 & 0.038 & -0.018\\
3-593.1 & 0.0018 & -0.018 & 0.018 \\
\end{tabular}
\caption{\label{tab:covar5}Covariance matrix for the $5.8\,\mu$m
  fluxes normalised to the maximum-likelihood values,
  i.e. Cov$(x_1,x_2)/(\mu_1\mu_2)$ where $x_1$ and $x_2$ are the
  variables being compared, and $\mu_1$ and $\mu_2$ are their maximum
  likelihood fluxes. The maximum-likelihood values
  themselves are given in table \ref{tab:photometry}.}
\end{table}

\begin{table}
\begin{tabular}{l|lll}
 & 3-586.0 & HDF850.1 & 3-593.1\\
\hline
3-586.0 & 0.0051 & -0.0064 & 0.0052\\
HDF850.1 & -0.0064 & 0.014 & -0.015\\
3-593.1 & 0.005 & -0.015 & 0.030 \\
\end{tabular}
\caption{\label{tab:covar8}Covariance matrix for the $8\,\mu$m
  fluxes normalised to the maximum-likelihood values,
  i.e. Cov$(x_1,x_2)/(\mu_1\mu_2)$ where $x_1$ and $x_2$ are the
  variables being compared, and $\mu_1$ and $\mu_2$ are their maximum
  likelihood fluxes. The maximum-likelihood values
  themselves are given in table \ref{tab:photometry}.}
\end{table}

\section{Modelling}\label{sec:modelling}
\subsection{Gravitational lens modelling}\label{sec:gravlens}
Both HDF850.1 and HDF8501.K have been argued to be only moderately
lensed by the foreground $z=1.224$ elliptical galaxy 3-586.0. In this
section we critically re-assess these claims.

We adopt a Singular Isothermal Ellipsoid (SIE) model for the
foreground lens, because in the image plane at the locations of
HDF850.1 and HDF850.1K, there are only very small differences between
the SIE model and a composite Navarro-Frenk-White profile plus de
Vaucouleurs' profile (e.g. Gavazzi et al. 2007) . The SIE model has a
critical radius 
$\theta_{\rm crit}$ defined as 
\begin{equation}
\left (\frac{\theta_{\rm crit}}{1{\rm ''}}\right ) = \left (\frac{\sigma}{186{\rm \,km\,s}^{-1}}
\right )^2 \left (\frac{D_{\rm LS}}{D_{\rm S}}\right )
\end{equation}
where $D_{\rm LS}$ and $D_{\rm S}$ are the angular diameter distances
between lens and source and between observer and source respectively,
and $\sigma$ is the velocity dispersion. In our adopted cosmology, the
source redshift of $z_{\rm s}=5.183$ and lens redshift $z_{\rm
  l}=1.224$ imply $D_{\rm S}=1238.3$\,Mpc and $D_{\rm
  LS}=638.0$\,Mpc. The model also has an
ellipticity, the magnitude and direction of which is determined by the
{\it HST F814W}-band imaging (Dunlop et al. 2004) to be $e=0.22$,
position angle $44.2^\circ$ East of North. 

To constrain the velocity dispersion $\sigma$, we use two
approaches. Firstly, di Serego Alighieri et al. (2005) present
rest-frame $B$-band fundamental plane observations of 18 $z\simeq1$
early-type galaxies from the K20 survey. Applied to the {\it HST
  F814W} data on 3-586.0, their best fit model predicts
$\sigma=129$\,km\,s$^{-1}$, and the dispersion of their data around
their best fit model implies an uncertainty in this prediction of $\pm
39$\,km\,s$^{-1}$. Secondly, Holden et al. (2005) present observations
of four early-type galaxies at $z=1.237$, almost exactly the same
redshift as our lensing galaxy 3-586.0. Scaling from each of these
galaxies assuming $\sigma^{1.2}\propto r_{\rm eff}I_{\rm eff}^{0.83}$
(where $r_{\rm eff}$ and $I_{\rm eff}$ are the effective radius and
surface brightness respectively), we obtain estimates of $\sigma$
ranging from $90$ to $253$\,km\,s$^{-1}$\,Mpc$^{-1}$, with a mean
$168\pm40$\,km\,s$^{-1}$\,Mpc$^{-1}$. Combining these estimates, we
obtain $\sigma=148\pm27$\,km\,s$^{-1}$\,Mpc$^{-1}$, implying
$\theta_{\rm crit}=0.34\pm0.12$ arcsec.

The magnification factors of HDF850.1 and HDF850.1K depend strongly on
$\theta_{\rm crit}$. Dunlop et al. (2004) argued that the lack of an
obvious counterimage to HDF850.1K implies this source must not be far
into the strong lensing regime. Neglecting the ellipticity, the
singular isothermal sphere (SIS) model predicts an image:counterimage
ratio of $(2\theta_{\rm crit}/\theta)-1$ for an image-counterimage
separation of $\theta$, and a total magnification $2\theta_{\rm
  crit}/(\theta-\theta_{\rm crit})$. This was used to argue that
$\theta_{\rm crit}<0.35$'' and that the total magnification of
HDF850.1K is $\mu<3.4$. However, the lack of a counterimage can simply
be due to the limited sensitivity of the $K$-band detection. Using the
{\sc gravlens} package (Keeton 2001) we found many SIE configurations
consistent with the data that violate these apparent SIS
constraints. In particular, an SIE model with $\theta_{\rm
  crit}=0.42$'' 
places a counterimage within $0.1$'' of the
location of a low signal-to-noise feature in the $K$-band
lens-subtracted image. In this model the magnification of HDF850.1K
would be $\mu=4.8$. The same model predicts the submm source HDF850.1 to have a
magnification of $\mu\simeq1.9\pm0.3$: at the location of the blue-shifted
peak of the [C{\sc ii}] emission in Walter et al. (2012) the
magnification is $1.61$, while at the red-shifted peak $\mu=2.12$.

\subsection{SED modelling}\label{sec:seds}

We performed SED fitting analysis on the detected HST sources 3-586.0
and 3-593.1 (Williams et al. 1996) as a check of the 
HDF850.1 and HDF850.1K photometry. To perform the
multi-wavelength fitting we adopted the popular \texttt{LePhare}
package (Arnouts et al. 1999; Ilbert et al. 2006). We use the
photometry available from various authors: F300W, F450W, F606W, F814W
WFCP2 photometry listed in Fernandez-Soto et al. (1999) and the WIRCam
Ks, IRAC 3.6, 4.5, 5.8, 8.0\,$\mu$m photometry reported by Wang et
al. (2010). As JH-WFC3 input fluxes we used the values estimated in
section \ref{sec:spitzer}. We then extrapolate from the best fit SED
the modelled photometry for each source, convolving our best fit model
in each case with the respective filter transmission curve.

These sources have known estimates of redshift:
3-593.1 lies at a photometric redshift of $z=1.76$ reported by
Fernandez-Soto et al. (1998), and 3-856.0, the lens elliptical galaxy,
has a spectroscopic redshift of $z=1.224$ obtained by Barger et
al. (2008); we can thus perform the fit fixing the redshift.

To perform the fit we adopt the extinction law by Calzetti et
al. (2000) and we use two different sets of template SED to check the
reliability of our fits: the SWIRE templates by Polletta et
al. (2007), together with some slightly modified versions from
Gruppioni et al. (2010), and the SED obtained by the stellar
population synthesis model by Bruzual \& Charlot (2003).  
We found no significant differences between the flux
predictions obtained by these two procedures and in table
\ref{tab:flux_predictions} below we report the resulting fluxes
obtained using the SWIRE templates. The $\chi^2$ of our best fit SED
are low (less than 2).  To give an estimate of the uncertainties we
should add to our modelled fluxes, we compare our modelled results to
the ones measured and given as input, where available.  In particular
we focus our attention on the range of wavelength around the JH-WFC3
bands which we want to recover completely by our model. In table
\ref{tab:flux_predictions} we thus report the predicted and measured
fluxes for the bands of our interest. The difference between the
modelled and the measured values are always between 2-5$\%$ (depending
on the wavelength) and well matched the uncertainties of the input
photometry.  We thus estimate that this is the uncertainty we should
expect to be associated to our JH-WFC3 predicted fluxes.  Finally we
estimate that the uncertainties coming from the use of a photometric
redshift for 3-593.1 is negligible or anyway included in the 5$\%$
uncertainties already discussed.

With our photometry of the submm galaxy we are also now able to model
the multi-wavelength SED of HDF850.1 and thus make the first
constraint on its stellar mass. We use the photometry reported by
Cowie et al. 2009 to complete our dataset and we perform the fit
following the same approach described above. In the case of HDF850.1
we fit from the NIR to the radio bands and we extrapolate the optical
photometry from the best solution. The input bands we used are J, H
WFC3, IRAC 5.8 and 8.0 $\mu$m (this work), ISO 15\,$\mu$m (as reported
by Downes et al. 1999), MIPS 24, 70 and 160\,$\mu$m, SCUBA
450\,$\mu$m, SCUBA 850\,$\mu$m and IRAM 1.3 (as reported by Cowie et
al. 2009). The J, H, $15\,\mu$m, $70\,\mu$m and $160\,\mu$m data are
upper limits.

\begin{figure}
\begin{center}
   {\includegraphics[width=0.45\textwidth]{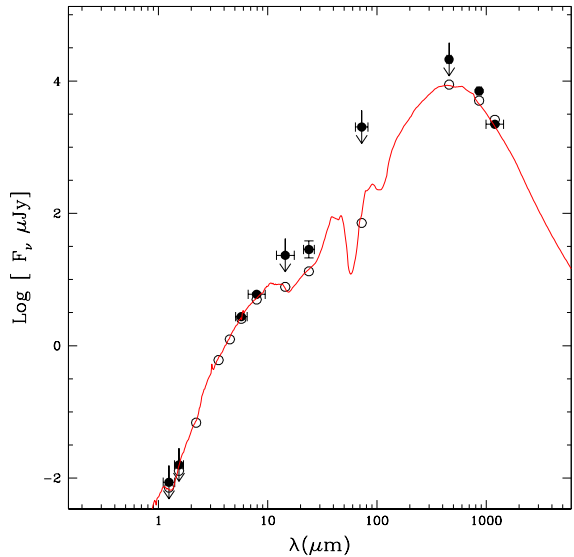}}
\end{center}
\caption{The best fit SED solution of HDF850.1. Solid red line: the
  best fit SWIRE SED solution corresponding to the IRAS\,20551-4250
  template from Polletta et al. (2007); full black circles: the input
  photometry as described in the text; open black circles: the model
  prediction at each band. The arrows stand for upper limits.}
\label{fig:fit}
\end{figure}

We tried to force the fit to reproduce the Arp 220 as this was the SED
claimed to be the best representation of the galaxy in Cowie et
al. 2009. We now know that the redshift of the galaxy is $z=5.183$
rather than the $z=4.1$ reported in Cowie et al., and we now find that
Arp 220 is not able to reproduce the photometry of the
galaxy. Instead, we find the best-fit SWIRE template is the
ultraluminous starburst galaxy IRAS\,20551-4250, shown in
Fig.\,\ref{fig:fit}.  The bolometric luminosity from $8-1000\,\mu$m is
$1.0\times10^{13}L_\odot$ (in agreement with the estimate reported by
Neri et al. 2014), implying a star formation rate of
$1700M_\odot$/year, with our assumed Salpeter initial mass function
from $0.1-100M_\odot$ before magnification correction and using the
conversion in Kennicutt 1998 (note that this is a factor of two higher
than the estimate in Walter et al. 2012).

To estimate the stellar mass using the SWIRE template SED of
IRAS\,20551-4250, we calculated the ratio of $J$-band rest-frame
luminosities of HDF850.1 and this local galaxy, and scaled the stellar
mass estimate from Mineo et al. (2012) to obtain a stellar mass
estimate of $3.4\times10^{11}M_\odot$, assuming a Salpeter initial
mass function from $0.1-100M_\odot$ (noting that the stellar
populations of local galaxies are older than would be found at
$z=5.2$).  
Alternatively, with the Bruzual \& Charlot model, with a Salpeter
initial mass function from $0.1-100M_\odot$ (converting the result
from Chabrier to Salpeter IMF using the conversion factor  described
by: $\log M_{\rm Chabrier} =  \log M_{\rm Salpeter} -  0.24$ as
reported in Santini et al. 2014) and a Calzetti extinction law with
$E(B-V)=0.5$, fitting to the $8\,\mu$m data and below, we obtain a
best fit stellar mass of $1.78\times10^{11}\,M_\odot$ and an age of
$1.02\times10^9$\,yr. The star formation history of our best fit
solution is described by a single stellar population with
exponentially decreasing star formation rate and an e-folding
timescale of $\tau=0.3\,$Gyr (star formation rate
$\propto\exp(-t/\tau)$). Such an age would imply an unphysically high
formation redshift; however, there is a degeneracy between the
reddening and the age. For $E(B-V)=0.75$ we obtain an age of
$0.6\times10^9$\,yr (formation redshift $z\simeq10$), and a stellar
mass of $2.75\times10^{11}\,M_\odot$. Combining all these estimates,
and treating the variation between them as a estimate of the
uncertainty (random or systematic), we conservatively quote a stellar
mass of $(2.5\pm1)\times10^{11}M_\odot$ before magnification
correction. 

\begin{table*}
\centering
\begin{tabular}{c|c|c|c|c|c|c|c|c}
\hline
{Source} & {Redshift} & {J/WFC3} & {H/WFC3} & {Ks/WIRCAM}  & {3.6 $\mu$m} & {4.5 $\mu$m} & {5.8 $\mu$m} & {8.0 $\mu$m}\\
\hline
3-586.0 & 1.22 &  21.24 / 21.35 & 20.34 / 20.60 & 22.87 / 23.15 & 18.80* / 13.83 & 16.07* / 11.68 & 10.32* / 7.39 &  5.62* / 5.34 \\
3-593.1 & 1.76 &  23.62 / 24.16 &  23.13 / 23.61 & 26.16 / 26.28 & 0.97* / 1.04 & 1.70* / 1.64 & 4.53* / 2.56 & 8.45* / 4.49 \\
\hline
\end{tabular}
\caption{\label{tab:flux_predictions} Measured / Predicted 
  JH-WFC3, Ks-WIRCAM total Vega magnitudes and IRAC 1234 total fluxes
  from the best fit SED of each galaxy. The fluxes are expressed in
  $\mu$Jy. The fluxes reported with * are considered as upper limits in the SED fitting.}
\label{fluxes}
\end{table*}

\section{Discussion: HDF850.1 in galaxy scaling relations}\label{sec:discussion}

A ``main sequence'' that has gained a great deal of 
popularity is the relationship between total stellar mass $M_*$ and
integrated star formation rate\footnote{The ``main
  sequence'' terminology in this and other galaxy scaling
  relationships discussed in this paper invites comparisons with the
  stellar main sequence in the Hertzsprung-Russell diagram, but note
  there is no suggestion that galaxies are homologous systems in
  general, nor that the astrophysics is as well-constrained.}
 $\dot{\rho}_*$ (e.g. Brinchmann et al. 2004, Daddi
et al. 2007, Elbaz et al. 2007, Salim et al. 2007, Zheng et al. 2007,
Noeske et al. 2007a,b, Pannella et al. 2009, Daddi et al. 2009, Stark
et al. 2009, Peng et al. 2010b, Gonz\'{a}lez et al. 2010, Rodighiero
et al. 2010, Karim et al. 2011, Rodighiero et al. 2011, Lilly et
al. 2013), often expressed in terms of the specific star formation
rate $\dot{\rho}_*/M_*$ that is observed to depend only weakly
on $M_*$ at a fixed epoch. The total stellar mass is necessarily
proportional to the average star formation rate of the galaxy (or its
progenitor systems) throughout the age of the Universe at that epoch,
so this relationship expresses whether the currently-observed star
formation rate is typical of the mean average history of the galaxy.

The age of the Universe at $z=5.183$ is only $1.1$\,Gyr, and the
specific star formation rate of HDF850.1 is $\dot{\rho}_*/M_*\simeq
8.5$\,Gyr$^{-1}$, so HDF850.1 is forming stars at a
rate approximately a factor of nine faster than its average up
to that point. Rest-frame UV-selected galaxies at this redshift have
specific star formation rates in the range $1-4$\,Gyr$^{-1}$ (Stark et
al. 2009, Bouwens et al. 2012, Gonz\'{a}lez et al. 2012), so HDF850.1
is also forming stars faster than coeval systems. The galaxies in
these samples are also typically less massive ($10^{7-10}M_\odot$), so
either HDF850.1 is not representative of lower-mass systems, or star
formation at these redshifts is usually episodic (e.g. Stark
et al. 2009). 

The proximity of HDF850.1K may suggest that a galaxy-galaxy
interaction may play a role in HDF850.1's high specific star formation
rate. If HDF850.1K were at the same redshift as HDF850.1, then the
physical separation between HDF850.1K and the submm galaxy would only
be $2.2\,$kpc, and the pair would be significantly differentially
magnified by the $z=1.22$ elliptical (e.g. Serjeant 2012). However,
our constraint on the $H-K$ colour of HDF850.1K of $H-K>3.8$ strongly
suggests that the $K$-band flux is either contaminated by the
foreground lens, or the detection itself is a spurious artefact of the
delicate PSF subtraction.

Elbaz et al. (2011) proposed another ``main sequence'' of galaxies in
the polycyclic aromatic hydrocarbon (PAH) luminosity versus bolometric
luminosity plane. High-$z$ starbursts curiously resemble scaled-up
versions of local star-forming disc galaxies, unlike local ULIRGs
which are under-luminous in PAH emission for their bolometric
luminosities.  Like CO $J=1-0$, PAH emission is assumed to be confined
to surface layers of GMCs. Variations in the amount of UV shielding by
dust are predicted to be responsible for most of the variation in CO
abundance in GMCs (e.g. Glover et al. 2010). It may be that a better
understanding of the turbulent conditions in the high-$z$ ISM will
also shed light on PAH-bolometric luminosity relationship, as well as
the Kennicutt-Schmidt and Elmegreen-Silk relations (see below). It
remains to be determined whether the lower PAH abundance in local
ULIRGs compared to higher-$z$ ULIRGs is because a larger fraction of
PAHs are exposed to hard radiation environments at low-$z$
(e.g. Guhathakurta \& Draine 1989), or e.g. because high-$z$
starbursts have more effective PAH replenishment from grain-grain
collisions (e.g. Rafikov 2006, Asano et al. 2013) or shocks
(e.g. Umana et al. 2010). As with H$_2$ and other ISM diagnostics,
direct detection of PAH emission in HDF850.1 and other $z>5$ galaxies
will have to wait for SPICA or JWST.

HDF850.1 has a molecular gas surface density of $\Sigma_{\rm
  gas}=1.4\times 10^9 M_\odot$\,kpc$^{-2}$ (from CO $J=5-4$, and
assuming a CO to H$_2$ conversion of $\alpha_{\rm
  CO}=0.8M_\odot$\,(K\,km\,s$^{-1}$\,pc$^{2}$)$^{-1}$), and a star
formation surface density of $\Sigma_{\rm SF}=35\,
M_\odot$\,yr$^{-1}$\,kpc$^{-2}$ (Walter et al. 2012), making it
typical of high-$z$ merger-driven starbursts on the Kennicutt-Schmidt
($\Sigma_{\rm gas}-\Sigma_{\rm SF}$) relation (e.g. Genzel et
al. 2010). The fact that the location on this relation appears to be
related to specific star formation rate has been used to argue for two
discrete modes of star formation in galaxies. 

However, before reading too much into the implications for HDF850.1,
it is worth reflecting on the physical mechanisms underpinning this
relation.  The Kennicutt-Schmidt relation is often interpreted in
terms of virialised molecular clouds that have, if not a single
characterisable size, then at least a well-characterisable average
size. In starburst galaxies and major merger systems the assumption of
virial equilibrium fails, so one uses a different CO to H$_2$
conversion. There has been considerable debate as to whether the
apparent bimodality in the Kennicutt-Schmidt relation has its origin
entirely in the assumed bimodal CO:H$_2$ conversion (e.g. Genzel et
al. 2010, Ivison et al. 2011, Narayanan et al. 2012, Sargent et
al. 2013). An alternative to the Kennicutt-Schmidt relation is the
Elmgreen-Silk relation, which uses the local dynamical timescale
$\tau_{\rm dyn}$ to relate $\Sigma_{\rm SF}$ to $\Sigma_{\rm
  gas}/\tau_{\rm dyn}$. The physical interpretation of this
relationship, unlike its Kennicutt-Schmidt counterpart, is to try to
characterise how large-scale dynamical processes partly govern the
star formation process (e.g. Elmegreen 1997, Silk 1997, Kennicutt \&
Evans 2012). Again, interpretations often invoke virialised or
marginally-bound molecular clouds (e.g. Silk 1997). However, numerical
simulations of GMCs in disc galaxies in no way resemble virialised
systems. Dobbs \& Pringle (2013) find the fraction of
gravitationally-bound GMCs to be $<20\%$ in their simulations; the
majority of molecular clouds are unbound. Glover et al. (2010) find
the CO and H$_2$ column densities in GMCs vary by at least two orders
of magnitude, and that CO gas abundance does not trace extinction,
though H$_2$ does at least trace the underlying gas distribution. In
both sets of simulations there is no clear distinction between
``clump'' and ``inter-clump'' mediums; much of the integrated CO
luminosity of a galaxy could lie in diffuse, low surface brightness
regions. It is perhaps only our own optical prejudices that lead us to
regard high-$A_{\rm V}$ regions as distinct entities at all.

The fact that these debates exist even at redshifts $z<2$ underlines
the difficulties in interpreting the $z>5$ population on these ``main
sequence'' relations, where both observations and modelling are less
well constrained. For example, the dependence of $\alpha_{\rm CO}$ on
metallicity has only been determined at $z<2.5$ (Mannucci et
al. 2010), and the metallicity of HDF850.1 has not been measured
directly. The interpretation of HDF850.1's location in these global
galaxy scaling relations, as with that of any of the highest-$z$
starbursts, must therefore await larger, higher-resolution numerical
models to provide better insights into the consequences of
observational constraints. The observational situation will improve
markedly with direct detections of redshifted H$_2$ with the SAFARI
instrument on SPICA (Roelfsema et al. 2012), and the H$_2$ would be
spatially resolved at sub-kpc scales in high-$z$ starbursts with the
proposed FIRI mission (Helmich, Ivison et al. 2007).  Spatially
resolving the gas and star formation on sub-kpc scales will be
strongly diagnostic of the physical processes (e.g. Hopkins, Narayanan
\& Murray 2013).

\section{Conclusions}\label{sec:conclusions}
The stellar mass of HDF850.1 is $(2.5\pm1)\mu^{-1}\times10^{11}M_\odot$
with a lensing magnification of $\mu\simeq1.9\pm0.3$, implying one of
the most extreme specific star formation rates in the galaxy ``main
sequence'' at $z>5$. The proposed HDF850.1K source is only $2.2\,$kpc
from the submm galaxy in projection, but our $H-K>3.8$ constraint on
the former suggests either its $K$-band flux is significantly
contaminated by the lensing galaxy or the $K$-band detection itself is
an artefact of the delicate subtraction of the foreground lens.

\section*{Acknowledgements}
The authors would like to thank the 
anonymous referee for helpful comments, and the 
Science and Technology Facilities
Council for support under grant ST/J001597/1. This work is based in
part on observations made with the NASA/ESA Hubble Space Telescope,
obtained from the data archive at the Space Telescope Science
Institute. STScI is operated by the Association of Universities for
Research in Astronomy, Inc. under NASA contract NAS 5-26555. This work
is also based in part on observations made with the {\it Spitzer} Space
Telescope, which is operated by the Jet Propulsion Laboratory,
California Institute of Technology under a contract with NASA.


\begin{thebibliography}{}

\bibitem{} Alexander, D.M., et al., 2003, AJ, 126, 539, 
\bibitem{} Arnouts, S., et al., 1999, MNRAS, 310, 540
\bibitem{} Asano, R.S., Takeuchi, T.T., Hirashita, H., Nozawa, T.,
  2013, arXiv:1303.5528
\bibitem{} Aussel, H., Cesarsky, C.J., Elbaz, D., Starck, J.L., 1999,
  A\&A, 342, 313
\bibitem{} Barger, A.J., et al., 1998, Nature, 394, 248
\bibitem{} Barger, A.J., Cowie, L.L., Wang, W.-H., 2008, ApJ, 689, 687 
\bibitem{} Bouwens, R.J., et al., 2012, ApJ, 753, 83
\bibitem{} Brandt, W.N., et al., 2001, AJ, 122, 2810
\bibitem{} Brinchmann, J., Charlot, S., White, S.D.M., et al., 2004,
  MNRAS, 351, 1151
\bibitem{} Bruzual, G., Charlot, S., 2003, MNRAS, 344, 1000
\bibitem{} Calzetti, D., et al., 2000, ApJ, 533, 682
\bibitem{} Capak, P., et al., 2004, AJ, 127, 180
\bibitem{} Combes, F., Rex, M., Rawle, T.~D., et al.\ 2012, A\&A, 538, L4 
\bibitem{} Conselice, C., et al., 2011, MNRAS, 413, 80
\bibitem{} Cowie, L.L., Barger, A.J., Wang, W.-H., Williams, J.P.,
  2009, ApJ, 697, L122
\bibitem{} Daddi, E., et al., 2007, ApJ, 670, 156
\bibitem{} Daddi, E., et al., 2009, ApJ, 694, 1517
\bibitem{} Dickinson, M., Giavalisco, M., et al., 2003, Proceedings of
  the ESO Workshop held in Venice, Italy, 24-26 October 2001;
  eds. R. Bender \& A. Renzini, p. 324.
\bibitem{} di Serego Alighieri, S., et al. 2005, A\&A, 442, 125
\bibitem{} Dobbs, C.L., Pringle, J.E., 2013, arXiv:1303.4995
\bibitem[Downes et al.(1999)]{1999A&A...347..809D} Downes, D., Neri, R., Greve, A., et al.\ 1999, A\&A, 347, 809 
\bibitem{} Dunlop, J.S., et al., 2004, MNRAS, 350, 769
\bibitem{} Eales, S.A., et al., 2010, PASP, 122, 499
\bibitem{} Elbaz, D., et al., 2007, A\&A, 468, 33
\bibitem{} Elbaz, D., et al., 2011, A\&A, 533, A119
\bibitem{} Elmegreen, B.G., 1997, Rev. Mex. A.A. (Serie de
  Conferencias), 6, 165
\bibitem{} Fernandez-Soto, A., Lanzetta, K.M., Yahil, A., 1999,
  ApJ, 513, 34
\bibitem{} Frayer, D.T., et al., 2011, ApJL, 726, 22
\bibitem{} Garrett, M.A., de Bruyn, A.G., Giroletti, M., Baan, W.A.,
  Schilizzi, R.T., 2000, A\&A, 361, L41
\bibitem{} Gavazzi, R., et al., 2007, ApJ, 667, 176
\bibitem{} Gavazzi, R., et al., 2008, ApJ, 677, 1046
\bibitem{} Genzel, R., et al., 2010, MNRAS, 407, 2091
\bibitem{} Glover, S.C.O., Federrath, C., Mac Low, M.-M., Klessen,
  R.S., 2010, MNRAS, 404, 2
\bibitem{} Gonz\'{a}lez, V., et al., 2010, ApJ, 713, 115
\bibitem{} Gonz\'{a}lez, V., et al., 2012, arXiv:1208.4362
\bibitem{} Griffin, M.J., et al., 2010, A\&A, 518, L3
\bibitem{} Grogin, N.A., et al., 2011, ApJS, 197, 35
\bibitem{} Gruppioni, C., Pozzi, F., Andreani, P., et al.\ 2010, A\&A, 518, L27 
\bibitem{} Guhathakurta, P., \& Draine, B.T. 1989, ApJ, 345, 230
\bibitem{} Guidetti, D., et al., 2013, MNRAS, 432, 2798
\bibitem{} Heiderman, A. Evans, N.J. II, Allen, L.E., Huard, T.,
  Heyer, M., 2010, ApJ, 723, 1019
\bibitem{} Helmich, F.P., Ivison, R.J., et al., 2007, arxiv:0707.1822
\bibitem{} Hogg, D.W., et al., 1997, AJ, 113, 474
\bibitem{} Holden, B.P., et al., 2005, ApJ, 620, L83
\bibitem{} Hopkins, P.F., Narayanan, D., Murray, N., 2013,
  arXiv:1303.0285 
\bibitem{} Hughes, D.H., et al., 1998, Nature 394, 241
\bibitem{} Ilbert, O., et al., 2006, A\&A, 457, 841
\bibitem{} Ivison, R.J., et al., 2011, MNRAS, 412, 1913
\bibitem{} Karim, A., et al., 2011, ApJ, 730, 61
\bibitem{} Kennicutt, R.C., 1998, ApJ, 498, 541
\bibitem{} Kennicutt, R.C., Evans, N.J. II, 2012, ARA\&A, 50, 531
\bibitem{} Koekemoer, A.M., ApJS, 197, 36K=
\bibitem{} Krist, J.E., Hook, R.N., Stoehr,F., 2011, Proc. SPIE, 8127,
  81270J
\bibitem{} Lilly, S.J., Carollo, C.M., Pipino, A., Renzini, A., Peng,
  Y., 2013, ApJ, in press (arXiv:1303.5059)
\bibitem{} Lupu, R.E., et al. 2012, ApJ, 757, 135
\bibitem{} Mannucci, F., Cresci, G., Maiolino, R. Marconi, A.,
  Gnerucci, A., 2010, MNRAS, 408, 1195
\bibitem{} Mineo, S., Gilfanov, M., Sunyaev, R., 2012, MNRAS submitted (arXiv:1207.2157)
\bibitem{} Morrison, G.E., Owen, F.N., Dickinson, M., Ivison, R.J.,
  Ibar, E., 2010, ApJS, 188, 178
\bibitem{} Murray, N., 2011, ApJ, 729, 133
\bibitem{} Narayanan, D., Bothwell, M., Dav\'{e}, R., 2012, MNRAS, 426,
  1178
\bibitem{} Negrello, M., et al., 2010, Science, 330, 800
\bibitem[Neri et al.(2014)]{2014arXiv1401.2396N} Neri, R., Downes, D., Cox, 
P., \& Walter, F.\ 2014, A\&A, in press (arXiv:1401.2396)
\bibitem{} Noeske, K.G., et al., 2007a, ApJ, 660, L43
\bibitem{} Noeske, K.G., et al., 2007b, ApJ, 660, L47
\bibitem{} Oliver, S.J., et al., 2012, MNRAS, 424, 1614
\bibitem{} Pannella, M., et al., 2009, ApJ, 698, L116
\bibitem{} Peng, C.Y., Ho, L.C., Impey, C.D., Rix, H.-W., 2010a, AJ,
  139, 2097 
\bibitem{} Peng, Y., et al., 2010b, ApJ, 721, 193
\bibitem{} Pineda, J.L., Goldsmith, P.F., Chapman, N., Snell, R.L.,
  Li, D., Cambr\'{e}sy, L., Brunt, C., 2010, ApJ, 721, 686
\bibitem{} Polletta, M., et al. 2007, ApJ, 663, 81
\bibitem{} Pope, A., \& Chary, R.-R., 2010, ApJ, 715, L171
\bibitem{} Rafikov, R.R., 2006, ApJ, 646, 288
\bibitem{} Riechers, D.~A., Bradford, C.~M., Clements, D.~L., et al.\ 2013, Nature, 496, 329 
\bibitem{} Rodighiero, G., et al., 2010, A\&A, 518, L25
\bibitem{} Rodighiero, G., et al., 2011, ApJ, 739, L40 
\bibitem{} Salim, S., et al., 2007, ApJS, 173, 267
\bibitem[Santini et al.(2014)]{2014A&A...562A..30S} Santini, P.,
  Maiolino, R., Magnelli, B., et al.\ 2014, A\&A, 562, A30 
\bibitem{} Sargent, M.T., et al., 2013, ApJ submitted (arXiv:1303.4392)
\bibitem{} Serjeant S., et al., 1997, MNRAS, 289, 457
\bibitem{} Serjeant, S., 2012, MNRAS, 424, 2429
\bibitem{} Silk, J., 1997, ApJ, 481, 703
\bibitem{} Smail, I., et al., 1997, ApJL, 490, L5
\bibitem{} Solomon, P.M., Rivolo, A.R., Barret, J., Yahil, A., 1987,
  ApJ, 319, 730
\bibitem{} Stark, D.P., et al., 2009, ApJ, 697, 1493
\bibitem{} Swinbank, A.M., et al., 2010, Nature, 464, 733
\bibitem{} Teplitz, H.I., et al., 2011, AJ, 141, 1 
\bibitem{} Teplitz, H.I., et al., 2013, AJ, 146, 159
\bibitem{} Umana, G., Buemi, C.S., Trigillo, C., Leto, P., Hora, J.L.,
  2010, ApJ, 718, 1036
\bibitem{} Walter, F., et al., 2012, Nature, 486, 233
\bibitem{} Wang, W.-H., Cowie, L.~L., Barger, A.~J., Keenan, R.~C., \& Ting, H.-C.\ 2010, ApJS, 187, 251 
\bibitem{} Williams, R.E., et al., 1996, AJ, 112, 1335
\bibitem{} Zheng, X.Z., et al., 2007, ApJ, 661, L41

\end{thebibliography}
\end{document}